\documentclass[
preprint,
showpacs,preprintnumbers,
bibnotes,
 amsmath,amssymb,
 aps,
 pra,
superscriptaddress,
longbibliography,
]{revtex4-1}

\usepackage{amsmath}
\usepackage{amsfonts}
\usepackage{amssymb}
\usepackage{graphicx}
\usepackage{xspace}
\usepackage{color}
\usepackage{comment}
\usepackage[hidelinks]{hyperref}
\usepackage{dcolumn}
\usepackage{bm}
\usepackage{mathtools}
\DeclarePairedDelimiter{\abs}{\lvert}{\rvert}

\usepackage{comment}

\usepackage[detect-weight=true, detect-family=true]{siunitx}

\graphicspath{ }

\begin{document}

\title{
Radiative control of dark excitons at room temperature by nano-optical antenna-tip induced Purcell effect
}

\author{Kyoung-Duck Park}
\affiliation
{Department of Physics, Department of Chemistry, and JILA,\\ University of Colorado, Boulder, CO, 80309, USA}
\author{Tao Jiang}
\affiliation
{Department of Physics, Department of Chemistry, and JILA,\\ University of Colorado, Boulder, CO, 80309, USA}
\author{Genevieve Clark}
\affiliation
{Department of Physics, Department of Materials Science and Engineering, University of Washington, Seattle, Washington 98195, USA}
\author{Xiaodong Xu}
\affiliation
{Department of Physics, Department of Materials Science and Engineering, University of Washington, Seattle, Washington 98195, USA}
\author{Markus B. Raschke}
\affiliation
{Department of Physics, Department of Chemistry, and JILA,\\ University of Colorado, Boulder, CO, 80309, USA}
\affiliation
{Center for Experiments on Quantum Materials,\\ University of Colorado, Boulder, CO, 80309, USA}
\email{markus.raschke@colorado.edu}
\date{\today}

\begin{abstract}

\noindent 
\textbf{Excitons, Coulomb-bound electron and hole pairs, are elementary photo-excitations in semiconductors, that can couple directly to light through radiative relaxation.
In contrast to these bright excitons, dark excitons X$\rm{_D}$ with anti-parallel electron spin polarization exist, with generally forbidden radiative emission. 
Because of their associated long lifetimes, these dark excitons are appealing candidates for quantum computing and opto-electronic devices.
However, optical read-out and control of X$\rm{_D}$ states have remained a major challenge due to their decoupling from light. 
Here, we present a novel tip-enhanced nano-optical approach to precisely switch and programmably modulate the X$\rm{_D}$ emission even at room temperature. 
Using monolayer two-dimensional transition metal dichalcogenide (TMD) WSe$\rm{_2}$ on a gold film as model system, we demonstrate ${\sim} 6 \times 10^5$-fold enhancement in dark exciton photoluminescence quantum yield. 
This is achieved by the unique coupling of the nano-optical antenna-tip to the dark exciton \textit{out-of-plane} optical dipole moment, with an extraordinary Purcell factor of $\ge 2 \times 10^3$ of the tip-sample nano-cavity.
Compared to the necessity of cryogenic temperatures and high magnetic fields in conventional approaches, our work provides a new way to harness excitonic properties in low-dimensional semiconductors and new strategies for quantum opto-electronic devices.} 
\end{abstract}

\maketitle

\noindent The two-dimensional (2D) nature of monolayer (ML) transition metal dichalcogenides (TMDs) creates tightly bound excitons with strong Coulomb interaction and an extraordinarily large binding energy \cite{splendiani2010, mak2010atomically, he2014}. 
Associated anomalous excitonic properties and strong light-matter interaction suggest a new paradigm for a range of applications in optoelectronics \cite{mak2016, basov2016, tong2017, hao2016}. 

The broken inversion symmetry and strong spin-orbit coupling (SOC) in ML TMDs lead to spin- and energy-splitting in the conduction band \cite{kosmider2013, echeverry2016}.
Associated anti-parallel electron spin configuration gives rise to two distinct states of bright and dark excitons with orthogonal transition dipole orientation, combined with the holes in the higher lying valence band as demonstrated recently both theoretically \cite{echeverry2016, slobodeniuk2016} and experimentally at low temperature \cite{molas2017, zhang2016, zhou2017}.

As known from zero-dimensional quantum dot (QD) studies, dark excitons X$\rm{_D}$ have a long lifetime due to solely non-radiative decay channels and spin flip processes \cite{smolenski2015}.
This distinct nature of dark excitons in low-dimensional semiconductors has attracted much attention for potential applications as coherent two-level systems for quantum information processing \cite{poem2010}, or Bose-Einstein condensation (BEC) \cite{combescot2007}, yet a full-scale study is hampered by the \textit{out-of-plane} transition dipole moments making them difficult to access optically \cite{molas2017, zhang2016, zhou2017}.

In order to induce dark exciton emission in atomically thin TMDs, different approaches were demonstrated in recent low temperature photoluminescence (PL) studies. 
Following a procedure as established for dark exciton emission in QDs \cite{nirmal1995, smolenski2012}, tilting the spin direction by applying a strong external \textit{in-plane} magnetic field ($\geq$ 14 T) induces a weakly allowed \textit{in-plane} optical transition by the Zeeman effect \cite{molas2017, zhang2016}. 
Alternatively, by extracting the intrinsically weak \textit{out-of-plane} optical transition by exciting \textit{out-of-plane} polarized surface plasmon polaritons (SPP) in a TMD-plasmonic device, weak radiative X$\rm{_D}$ emission through dark exciton-SPP coupling can be induced \cite{zhou2017}.
However, even then the SPP-induced X$\rm{_D}$ emission is still restricted to cryogenic temperature conditions, due to the small energy difference between dark and bright exciton ensuring low thermal excitation into the otherwise overwhelming bright exciton emission channel.

In this work, we demonstrate a new approach for dark exciton spectroscopy based on state selective dark exciton coupling with tip-enhanced PL (TEPL) spectroscopy, as illustrated in Fig. 1a-b.
Here, the scanning probe nano-optical antenna-tip selectively couples to the \textit{out-of-plane} transition dipole moment which facilitates Purcell-enhanced dark exciton emission with the few-fs radiative dynamics \cite{kravtsov2014}.
With this simple and generalizable approach we demonstrate excitation, modulation, and radiative control of dark exciton emission, at room temperature, and with high quantum yield.
The combination of the nanoscale localized ($\leq$ 15 nm) effective excitation through strongly confined \textit{out-of-plane} optical fields at the tip-Au substrate nano-gap, with antenna-tip mediated near- to far-field mode transformation gives rise to a ${\sim}6 \times 10^5$-fold X$\rm{_D}$-PL enhancement as we demonstrate in WSe$\rm{_2}$ with a Purcell factor $\ge 2 \times 10^3$ enhanced spontaneous emission rate.
From precise atomic force microscopy (AFM) nano-gap distance control of the dark exciton to antenna-tip coupling strength, we achieve from simple switching to active modulation of the dark exciton \textit{On/Off} states in time and space.
\\
\begin{figure*}
	\includegraphics[width = 15 cm]{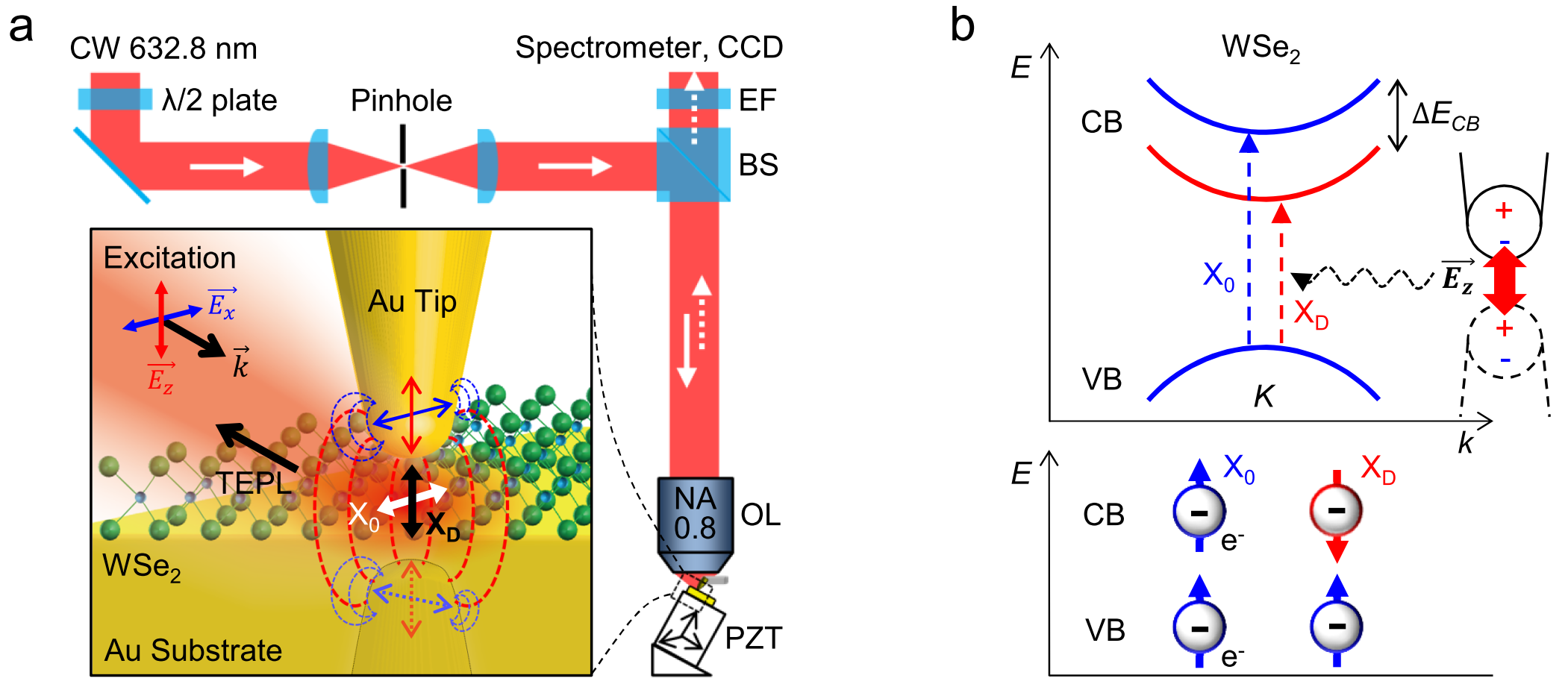}
	\caption{
\textbf{Schematic of tip-enhanced photoluminescence (TEPL) spectroscopy and electronic band structure of WSe$\rm{_2}$.} 
(a) Selective excitation and probing of the transition dipole moments of dark (\textit{out-of-plane}) and bright (\textit{in-plane}) excitons by polarization control. 
EF: edge filter, BS: beam splitter, OL: objective lens.
(b) Split-band and -spin configurations of bright and dark exciton states. 
Spin-forbidden optical transition of dark excitons is induced by strongly confined local antenna fields with plasmonic Purcell enhancement at the tip-sample nano-gap.
CB: conduction band, VB: valence band.
}
	\label{fig:setup}
\end{figure*}

\noindent
{\bf Radiative emission of dark excitons at room temperature}

\noindent
The experiment is based on TEPL spectroscopy \cite{park2016tmd}, with side illumination of Au nano-tip manipulated in a shear-force AFM as shown schematically in Fig. 1a (see Methods for details).
The Au tip is oriented normal with respect to a planar Au substrate. 
TEPL spectroscopy is then performed by tip-sample distance control between the Au tip and a ML WSe$\rm{_2}$ transferred onto the Au substrate. 
All experiments are performed at room temperature.
 
\begin{figure*}
	\includegraphics[width = 16 cm]{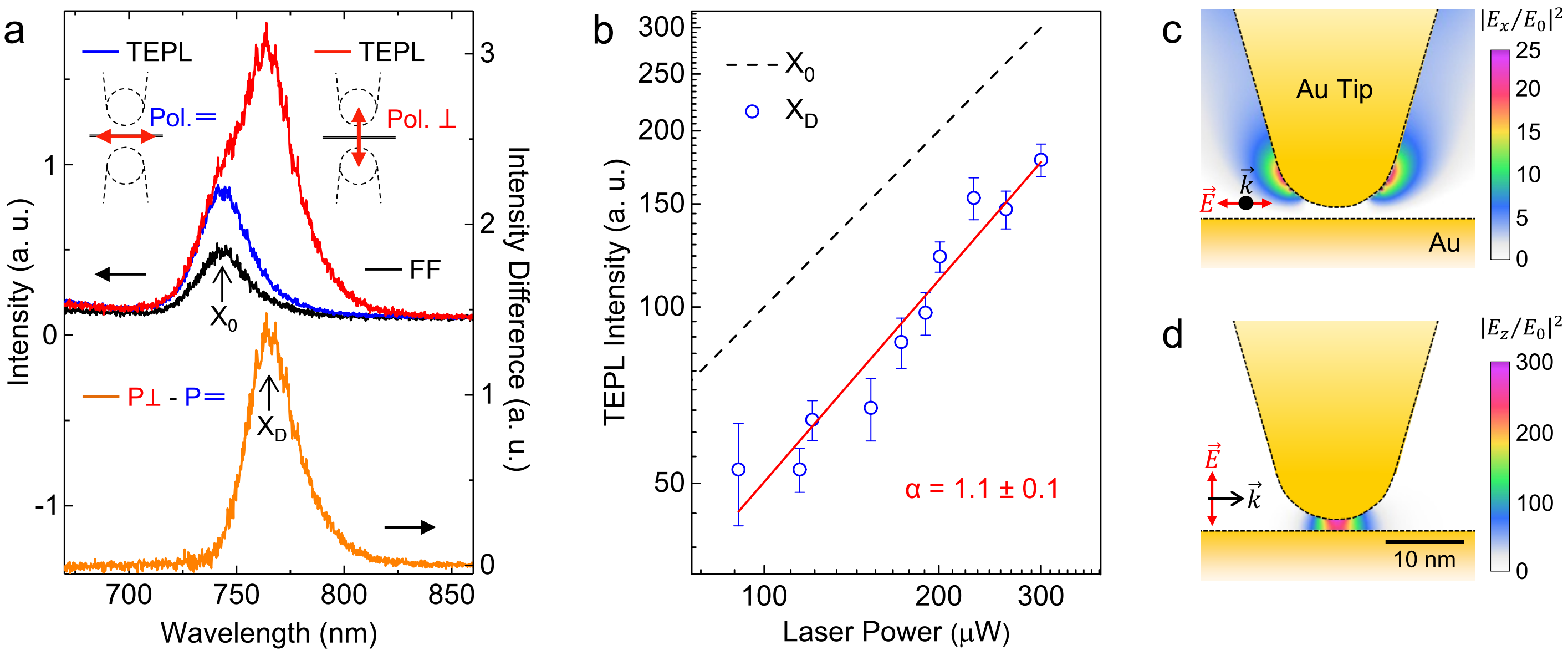}
	\caption{
\textbf{Probing radiative emission of dark excitons of WSe$\rm{_2}$ through polarization- and power-dependence of TEPL.} 
(a) Excitation polarization dependent TEPL spectra of monolayer WSe$\rm{_2}$ on a Au substrate at ${\sim}1$ nm tip-sample distance with tip-selective X$\rm{_D}$ emission (orange). 
(b) Log-plot of the power dependence of TEPL intensity of dark (X$\rm{_D}$) and bright (X$\rm{_0}$) exciton. 
(red line is a fit of the dark exciton emissions exhibiting a linear power dependence).
(c-d) Finite-difference time domain (FDTD) simulation of the \textit{in-plane} (c) and \textit{out-of-plane} (d) optical field intensity and confinement under experimental conditions of (a). 
}
	\label{fig:tepl}
\end{figure*}

Fig. 2a shows TEPL spectra at 1 nm tip-sample distance with excitation polarization oriented parallel or perpendicular with respect to the sample surface. 
A strong X$\rm{_D}$ emission peak is observed (red). 
In contrast, the X$\rm{_D}$ response is suppressed for tip-perpendicular polarization (blue) even though enhanced PL response of X$\rm{_0}$ is observed, attributed to an \textit{in-plane} localized optical field effect in agreement with the spectral X$_0$ characteristics in far-field emission (black).
The difference spectrum (orange) then corresponds to the pure X$\rm{_D}$ mode emission.
From Lorentzian fitting of the spectra for X$\rm{_D}$ and X$\rm{_0}$, we obtain ${\sim}46$~meV of intravalley energy splitting between the dark and bright excitons (see Supplementary Fig. S1). 
This energy difference of X$\rm{_D}$ peak position is in good agreement with recent X$\rm{_D}$ observations facilitated by \textit{in-plane} magnetic field (47 meV) \cite{zhang2016} and SPP coupling (42 meV) \cite{zhou2017}. 
Further the TEPL linewidth of the X$\rm{_D}$ emission is more narrow than the bright exciton linewidth, in agreement with previous observation (see Supplementary Information for details) \cite{zhou2017}.

To test for a possible contribution from bi-exciton emission, which has comparable photon energy to dark states, we measure the TEPL intensity as a function of the excitation power for X$\rm{_D}$ and X$\rm{_0}$ emissions.
Fig. 2b shows the resulting TEPL power dependence based on linear fits to the TEPL spectra and plotting the integrated spectral intensities of X$\rm{_D}$ and X$\rm{_0}$ emissions. 
On the basis of the linear power dependence behavior, we exclude bi-exciton emission \cite{you2015}.

To understand the polarization dependence of the X$\rm{_D}$ emission, and to model the intensity difference of \textit{in-plane} and \textit{out-of-plane} local fields at the Au tip - Au substrate junction, we calculate the confined optical field intensity using finite-difference time-domain (FDTD) simulation for our experimental conditions (Fig. 2a). 
As shown in Fig. 2c-d, in the nano-gap, the optical field intensity of the \textit{out-of-plane} mode is stronger by a factor of $\geq 3 \times 10^2$ than that of the \textit{in-plane} mode as expected.

\begin{figure*}
	\includegraphics[width = 16 cm]{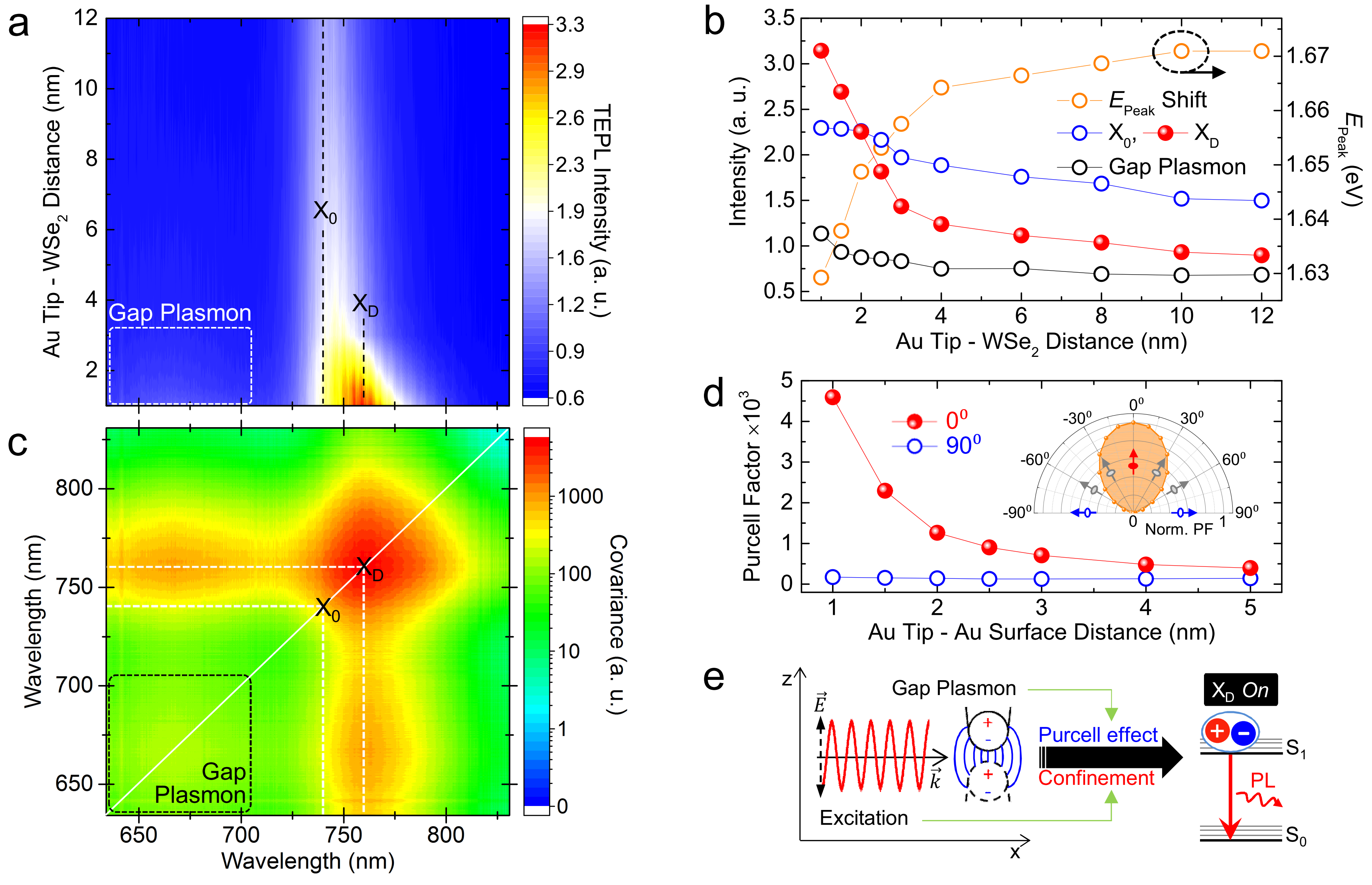}
	\caption{
\textbf{Active control of tip-induced radiative emission of dark excitons of WSe$\rm{_2}$ monolayer.} 
(a) Tip-sample distance dependence of TEPL spectra. 
(b) Tip-sample distance dependence of peak intensity of TEPL responses (X$\rm{_D}$, X$\rm{_0}$, and gap plasmon) with spectral energy shift, derived from (a). 
(c) 2D covariance map of the distance dependent TEPL spectra from (a), exhibiting strong correlation between the gap plasmon and X$\rm{_D}$ emission. 
(d) Simulated plasmonic Purcell factor for \textit{out-of-plane} and \textit{in-plane} spontaneous dipole emitters with respect to the distance between Au tip and Au surface.
Inset: Normalized plasmonic Purcell factor with respect to emitter orientation.
(e) Illustration of the tip-enhanced emission mechanism of dark excitons as a combination of both enhanced excitation (\textit{out-of-plane} optical field confinement) and antenna-mediated emission (polarization transfer and enhanced radiative decay by Purcell effect) at the plasmonic nano-gap.
}
	\label{fig:spec}
\end{figure*}

For further investigation of the antenna-tip induced dark exciton emission, we perform TEPL measurements under precise nanometer tip-sample distance control.
Fig. 3a shows contour plots of TEPL spectra of ML WSe$\rm{_2}$ with respect to the distance between the Au tip and Au substrate. 
Bright exciton emission (X$\rm{_0}$) at 743.5 nm (1.667 eV) is observed in the distance region of 4 - 12 nm attributed to the enhanced localized (\textit{in-plane}) near-field excitation at the tip apex. 
In contrast, for shorter distances in the 1 - 3 nm range, dark exciton X$\rm{_D}$ emission emerges and dominates the spectra at 765 nm (1.621 eV).
In addition, a weak tip-plasmon PL response from the Au - Au nano-gap is seen as described previously \cite{kravtsov2014}.
Fig. 3b shows corresponding distance dependence in TEPL intensity for X$\rm{_D}$, X$\rm{_0}$, and gap plasmon emission.
While the X$\rm{_0}$ peak intensity saturates below 2 nm distances due to polarization and energy transfer and non-radiative relaxation in the metal tip and substrate, the X$\rm{_D}$ peak intensity continues to rise sharply correlated with an increasing gap plasmon PL intensity.
Because the \textit{in-plane} dipole of the X$\rm{_0}$ does not couple to the antenna mode, the PL quenching dominates over enhancement at short distances. 
In contrast, the X$\rm{_D}$ emission with its \textit{out-of-plane} dipole, when coupled to the antenna mode with its fs-radiative decay, continues to dominate at short distances \cite{kravtsov2014, kuhn2006, anger2006}. 
 
This behavior is most clearly evident in the 2D covariance plot (Fig. 3c).
From the full data set of Fig. 3a, we calculate the covariance $\sigma(i,j)$ between wavelengths $i$ and $j$ from the distance dependent TEPL intensities $I(i,d)$ using
\begin{equation}
\label{Covariance}
\sigma{(i,j)}=\frac{1}{N}\sum\limits_{d=d_1}^{d_n}[I(i,d)-\left\langle{I}(i)\right\rangle]\times[I(j,d)-\left\langle{I}(j)\right\rangle].
\end{equation}  
The resulting 2D covariance map clearly shows strong (weak) correlation between the \textit{out-of-plane} gap plasmon and the dark (bright) exciton emission.
The plasmon PL emission serves as an indicator and metric of the X$\rm{_D}$-tip polarization transfer and rapid Purcell enhancement of PL emission as established previously (see Ref. \cite{kravtsov2014} and Discussion for details).
\\

\begin{figure}
	\includegraphics[width = 16 cm]{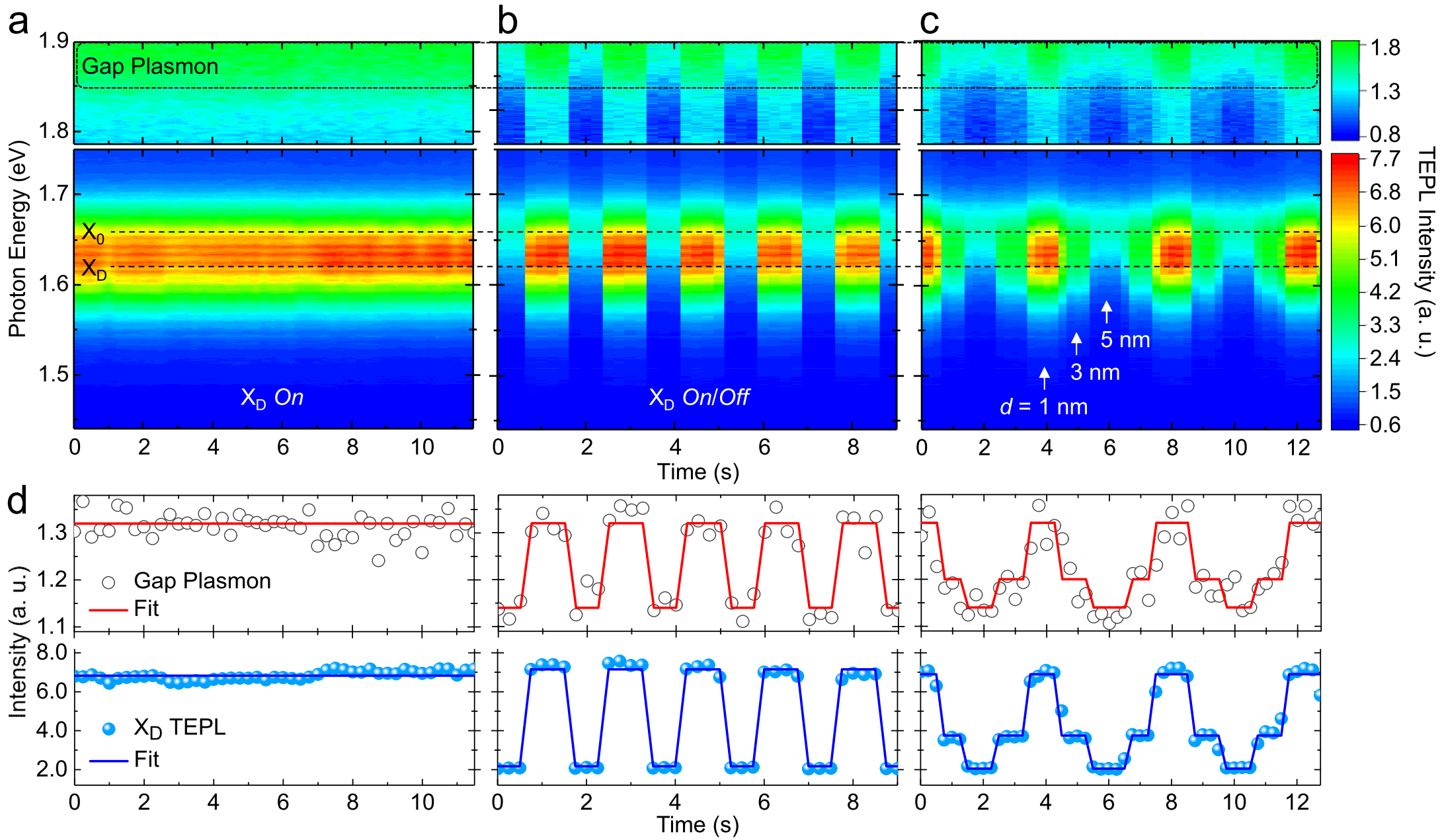}
	\caption{
\textbf{Switching and modulation of dark exciton emission.} 
(a) Time-series TEPL response (X$\rm{_D}$, X$\rm{_0}$, and gap plasmon PL) in continuous \textit{On} state of X$\rm{_D}$ emission at 1 nm tip-sample distance. 
(b) Discrete \textit{On/Off} switching of X$\rm{_D}$ emission with discrete tip-sample distance variation between 1 nm and 5 nm. 
(c) X$\rm{_D}$ emission modulation by functional control of tip-sample distance. 
(d) Time-series of peak intensity of TEPL response with correlation between corresponding X$\rm{_D}$ emission and gap plasmon, derived from (a)-(c).
}
	\label{fig:edge}
\end{figure}

\noindent
{\bf Radiative control of dark excitons}

\noindent
As we demonstrated previously \cite{park2016tmd}, the shear-force AFM tip can also act as an active control element to modify excitonic properties of TMDs both spectrally and spatially. 
In the optical antenna tip-WSe$\rm{_2}$-Au surface configuration, the nano-gap is regulated by a shear-force feedback mechanism \cite{park2010}. 
This gives rise to sub-nm precise control of the gap plasmon response, and associated strongly correlated dark exciton emission.
Fig. 4a-c show time-series contour plots of the TEPL response (X$\rm{_D}$, X$\rm{_0}$, and gap plasmon) during constant (a), discrete (b), and modulated (c) tip-sample distance. 
The precision of nano-gap control relies on the mechanical quality (Q) factor of the AFM tuning forks. 
We generally achieve Q-factors of $\geq 4 \times 10^3$ by attaching two Au tips to both prongs of the tuning fork. 
Using this high-Q sensor, we regulate a tip-sample distance with ${\sim}0.2$ nm precision, and stably maintain the X$\rm{_D}$ \textit{On} state under ambient conditions with minimal fluctuations (a). 
Varying the distance between 1 nm and 5 nm the X$\rm{_D}$ can be switched discretely between its \textit{On} and \textit{Off} states (b).
Correspondingly, by continuously time-varying the tip-sample distance, we can programmably modulate the X$\rm{_D}$ emission (c).    
Fig. 4d shows the derived time-series of peak intensity of TEPL response from Fig. 4a-c, with correlation of X$\rm{_D}$ PL and gap plasmon, verifying the precise control of the X$\rm{_D}$ emission. 
\\

\noindent
{\bf Radiative decay mechanisms of dark excitons}

\noindent 
In 2D TMDs, as dictated by the \textit{C$_{3h}$} point group and resulting selection rules, only an \textit{out-of-plane} optical transition is allowed with ${\sim}10^{-3}-10^{-2}$ times smaller radiative decay rate compared to an \textit{in-plane} optical transition dipole of the bright excitons \cite{echeverry2016, slobodeniuk2016}.

In principle, a spin flip is required to induce radiative decay of the dark excitons \cite{slobodeniuk2016}.
The electron spins can be extrinsically flipped by applying an \textit{in-plane} magnetic field \cite{zhang2016}.
Using a large field of $\geq$ 14 T, a radiative decay rate $\Gamma$ $\sim 10^{-3}\Gamma_B$ can be obtained with \textit{in-plane} optical transition by the Zeeman effect \cite{molas2017, zhang2016}, where $\Gamma_B$ is the radiative decay rate of the bright excitons. 
On the contrary, when the reflection symmetry in the surface normal direction is broken (typical condition for TMD crystals on a flat substrate), an intrinsic spin flip is facilitated by a virtual transition in the conduction band attributed to the SOC mixing. 
This then induces an \textit{out-of-plane} dipole transition by the perturbation of a local field \cite{bychkov1984, slobodeniuk2016}. 
Since this transition is between quantum states with identical magnetic quantum number and opposite parity, the transition dipole moments give rise to Bychkov-Rashba coupling to an \textit{out-of-plane} optical field \cite{ochoa2013}. 
The associated radiative decay rate $\Gamma_D$ $\sim (10^{-3}-10^{-2})\Gamma_B$ is estimated theoretically \cite{echeverry2016, slobodeniuk2016}, and is independent of an external field.

In our experimental design, a combination of two physical mechanisms of increase in excitation rate by field enhancement as well as Purcell factor induced optical antenna emission (Fig. 3e) is responsible for the ability to detect the dark exciton modes via tip-coupling to an \textit{out-of-plane} field.
First, a strongly confined \textit{out-of-plane} optical field effectively excites the transition dipole moment of dark excitons with $|\textit{E}_z|^2$ in the nano-gap enhanced by a factor of $\geq 3 \times 10^2$ compared to the incident field intensity $|\textit{E}_0|^2$ (Fig. 2d). 
In addition, and most significantly, the spontaneous emission rate is enhanced in the nano-gap due to the plasmonic Purcell effect \cite{akselrod2014, wang2016}.
Here, as demonstrated previously \cite{kravtsov2014}, near-field dipole-dipole coupling and exciton polarization transfer into the tip with its few-fs radiative lifetime of its plasmonic optical antenna mode, gives rise to an increased emission of the dark exciton with decreasing tip-sample distance.  
To model this effect we computationally design \textit{out-of-plane} and \textit{in-plane} fluorescent model dipole emitters positioned within the nano-gap to calculate the effective Purcell factor ($\gamma_{PF}$ = $\gamma$/$\gamma_0$, See Supplementary Information for details). 
As shown in Fig. 3d, the Purcell factor of an \textit{out-of-plane} spontaneous emitter in the tip-surface gap ($0^{\circ}$) exceeds $2 \times 10^3$ for $\leq{1.5}$ nm distances as a result of the polarization transfer and rapid radiative tip emission.
\\

\noindent
{\bf Enhancement factor of tip-induced dark exciton emission}

\noindent
For our experimental condition, the resulting overall TEPL intensity enhancement factor (EF) of the dark exciton emission
as a combination of both effects given by
\begin{equation}
\label{TED}
<\text{EF}> = \abs*{\frac{\textit{E}_z}{\textit{E}_0}}^2 \cdot \gamma_{PF}, 
\end{equation}
is estimated to be as high as ${\sim}6 \times 10^5$ at 1.5 nm plasmonic nano-gap.
This enhancement factor of the dark excitons is, to the best of our knowledge, the largest enhancement factor for fluorescent emitters investigated to date.
In comparison, even refined plasmonic Ag nanocubes coupled to a Ag film achieved ${\sim}3 \times 10^4$-fold fluorescence enhancement from cyanine molecules as the closest analogue \cite{rose2014, akselrod2014}. 
This extraordinary PL enhancement is due to the strictly \textit{out-of-plane} oriented transition dipole moment in the atomically thin semiconductor sandwiched in the $\leq$ 1.5 nm plasmonic nano-gap.

This \textit{out-of-plane} mode selective enhancement facilitates the room temperature observation of X$\rm{_D}$ not readily possible with other techniques. 
We can quantify the ratio of near-field TEPL intensity of the dark excitons ($I^{NF}_{D}$) compared to the far-field PL intensity of the bright excitons ($I^{FF}_B$) given by 
\begin{equation}
\label{TEDark}
\frac{I^{NF}_{D}}{I^{FF}_B} \approx \abs*{\frac{\textit{E}_z}{\textit{E}_0}}^2 \cdot \gamma_{PF} \cdot \frac{\Gamma_D}{\Gamma_B} \cdot \frac{\eta_D}{\eta_B}.
\end{equation}
Here, $\Gamma_D$ and $\Gamma_B$ represent respective decay rates, modified by the Purcell factor $\gamma_{PF}$, and $\eta_D$ and $\eta_B$ are relative occupation numbers of thermally populated dark and bright excitons, respectively, with their ratio $\eta_D$/$\eta_B$ given by $e^{\frac{\Delta E}{k_B T}}$ with $\Delta E$ the energy splitting between the dark and bright excitons \cite{crooker2003}.
With the parameters from above, $I^{NF}_{D}/{I^{FF}_B} \sim4 \times 10^3 - 4 \times 10^4$ at 300 K, thus facilitating direct probing and active control of the tip-enhanced radiative emission of the dark excitons with strong contrast even at room temperature.
However, in the absence of the antenna tip polarization and Purcell effect enhancement, the recently demonstrated dark exciton emission at low temperature by an \textit{in-plane} magnetic field \cite{molas2017, zhang2016} and SPP coupling \cite{zhou2017} does not readily extend to room temperature due to the dominant spectral weight of the bright excitons.
\\

\noindent
{\bf Conclusions}

\noindent
Our new approach thus gives access to potential applications of dark excitons in quantum nano-optoelectronics over a wide temperature range. We envision the demonstrated tip-antenna platform for room temperature dark exciton emission with or without nano-opto-mechnical control as ideal building block for functional quantum devices. 
Further, the nanoscale optical switching of the spin states paves way for new design and fabrication of nano-spintronic devices.
Specifically, the control of long-lived dark excitons confined in only ${\sim}150$ nm$^3$ mode volume can be exploited to create nanoscale devices for integrated quantum-photonic circuits and active quantum information processor such as nano-light emitting diodes, nano-optical switch/multiplexer, high-density memory, and qubit. 
The nano-confinement further allows for imaging with $\leq$ 15 nm spatial resolution of heterogeneity of dark excitonic properties in 2D TMDs \cite{park2016tmd}, with the possibility for an additional modulation in electronic energy with local strain engineering via nano-mechanical tip force control as we demonstrated recently \cite{park2016tmd}. 
The range of dynamic controls including coherent ultrafast excitation and tip/antenna manipulation thus gives access to a range of new phenomena at the sub-10 nm scale regime including room temperature strong coupling \cite{chikkaraddy2016, kleemann2017}, interlayer electron-phonon coupling \cite{jin2016}, or \textit{out-of-plane} exciton behaviors in single molecules and QDs beyond van der Waals materials \cite{rivera2016}.
\\

\noindent
{\bf Methods}

\noindent
\textbf{\textit{Sample preparation}}
WSe$_2$ monolayers are grown by physical vapor transport using powdered WSe$_2$ as precursor material.
Source material (30 mg) in an alumina crucible is placed in the hot zone of a 25.4 mm horizontal tube furnace, and an SiO$_2$ substrate is placed downstream in a cooler zone at the edge of the furnace ($750-850$~$^{\circ}$C). 
Before growth, the tube is evacuated to a base pressure of 0.13 mbar and purged several times with argon. 
The furnace is then heated to 970~$^{\circ}$C at a rate of 35~$^{\circ}$C/min and remains there for a duration of 5-10 min before cooling to room temperature naturally. 
A flow of 80 sccm argon and 20 sccm hydrogen is introduced as carrier gas during the 5-10 min growth period. 
The as-grown WSe$_2$ crystals are then transferred onto flat template stripped Au substrates using a wet transfer technique.
For that purpose, PMMA (6\% in anisole) is spin-coated onto the SiO$_2$ wafer, covering the region with WSe$_2$ monolayer crystals. 
The wafer is then placed in a solution of dilute hydrofluoric acid (20\% in distilled water), until the SiO$_2$ layer is etched away and the PMMA membrane with WSe$_2$ crystals floats free. 
The membrane is then rinsed in distilled water to remove residual etchant, and scooped up onto a wire loop.
The PMMA membrane can then be placed onto the Au sbstrates under an optical microscope, similar to the commonly used dry transfer technique. 
Once the membrane has been lowered into contact with the substrate, heating the substrate $\geq$ 160~$^{\circ}$C melts the PMMA layer releasing it from the wire loop. \newline
\noindent
\textbf{\textit{TEPL spectroscopy setup}}
In the TEPL spectroscopy setup, the sample is mounted to a piezoelectric transducer (PZT, P-611.3, Physik Instrumente) with sub-nm precision positioning.
Electrochemically etched Au tips (${\sim}10$~nm apex radius) are attached to a quartz tuning fork (resonance frequency = ${\sim} 32$~kHz) \cite{park2016tmd}. 
To regulate the tip-sample distance, the AFM shear-force amplitude is monitored and controlled from the electrically driven tuning fork \cite{karrai1995}. 
Coarse tip positioning is performed using a stepper motor (MX25, Mechonics AG), and shear-force feedback and sample position are controlled by a digital AFM controller (R9, RHK Technology).
The sample surface is tilted by ${\sim}60$$^{\circ}$ with respect to the incident k-vector for effective excitation.
A Helium-Neon laser beam (632.8~nm, $\leq$ 0.3~mW), after passing through a half wave plate for polarization control, is focused into the junction between the Au substrate and the tip apex by an objective lens (NA=0.8, LMPLFLN100$\times$, Olympus). 
TEPL signal is collected in backscattered direction, passed through an edge filter (633~nm cut-off) and detected using a spectrometer (f = 500 mm, SpectraPro 500i, Princeton Instruments) with a thermoelectrically cooled electron-multiplied charge-coupled device (CCD, ProEM+: 1600 eXcelon3, Princeton Instruments). 
The spectrometer is calibrated using hydrogen and mercury lines, and a 150 grooves/mm grating is used to obtain a high bandwidth spectrum for simultaneous measurement of gap plasmon and TEPL spectra.

\bibliography{tmd_2} 

\vskip 1cm
\noindent
{\bf Acknowledgements}

\noindent
The authors would like to thank Mikhail D. Lukin for insightful discussions. 
K.-D. Park, T. Jiang, and M. B. Raschke acknowledge funding from the U.S. Department of Energy, Office of Basic Sciences, Division of Material Sciences and Engineering, under Award No. DE-SC0008807.
G. Clack and X. Xu acknowledge support from NSF-EFRI-1433496.
We also acknowledge support provided by the Center for Experiments on Quantum Materials (CEQM) of the University of Colorado.
\\

\noindent
{\bf Author contributions}

\noindent
M.B.R. and K.-D.P. conceived the experiment.
K.-D.P. performed the measurements and the FDTD simulations.
K.-D.P. and M.B.R. designed the samples, and G.C. and X.X. prepared the samples.
K.-D.P. and M.B.R. analysed the data, and all authors discussed the results.
K.-D.P. and M.B.R. wrote the manuscript with contributions from all authors. 
M.B.R. supervised the project.
\\

\noindent
{\bf Competing financial interests}

\noindent
The authors declare no competing financial interests.

\end{document}